\documentclass[12pt,letterpaper]{article}

\usepackage{amsmath}
\usepackage{color}
\usepackage{amsfonts}
\usepackage{amssymb}
\usepackage{graphicx}
\usepackage{hyperref}

\def\be{\begin{equation}}
\def\ee{\end{equation}}

\begin{document}

\begin{center}
\Large{\textbf{The  Memory of Primordial Gravitational Waves}}
\\[0.5cm]
\large{Paolo Creminelli$^{\,\rm a, \rm b}$ \footnote{Email: {\tt creminel@ictp.it}} and Filippo Vernizzi$^{\,{\rm c}}$  \footnote{Email: {\tt filippo.vernizzi@ipht.fr}}}
\\[0.5cm]

\small{
\textit{$^{\rm a}$
ICTP, International Centre for Theoretical Physics\\ Strada Costiera 11, 34151, Trieste, Italy}}
\vspace{.2cm}

\small{
\textit{$^{\rm b}$
IFPU - Institute for Fundamental Physics of the Universe,\\ Via Beirut 2, 34014, Trieste, Italy }}
\vspace{.2cm}

\small{
\textit{$^{\rm c}$ Institut de Physique Th\'eorique, Universit\'e Paris Saclay, \\CEA, CNRS, 91191 Gif-sur-Yvette, France}}
\vspace{.5cm}

\vspace{0.5cm} 

\begin{abstract}\normalsize
Primordial gravitational waves, after they enter the horizon and decay away, leave a residual displacement  in test particles: a memory, in  analogy with gravitational waves generated by astrophysical sources. The late-time distance between test particles is related to the one at early times by $\xi^i_{\rm late} = \frac{a_{\rm late}}{a_{\rm early}} (\delta^i_j -\frac12 \bar h^i_j)\xi^j_{\rm early}$. Therefore, the deformation of an initial spherical shell does not depend on the cosmological evolution, but only on the primordial value $\bar h^i_{j}$ of the gravitational wave. 
 The memory is thus related to the adiabatic tensor mode that maps the unperturbed FLRW geometries at early and late times; this is analogous to the relation between memory in Minkowski spacetime and the BMS  group. The primordial memory is also connected to the consistency relations of cosmological correlators, as the flat-space memory is related to the soft theorems for gravitational wave emission. We comment on the signature of the effect on the CMB $B$-modes  and on the large-scale structure. There is also a primordial memory effect that is subleading in the spatial gradients of the wave: it is encoded in the rotation of free-falling gyroscopes.

\end{abstract}

\vspace{0.3cm}

{\em To the memory of Valery~A.~Rubakov. \\
Contribution to the special issue of the International Journal of Modern Physics A.}

\vspace{0.3cm}

\end{center}

\newpage

The memory effect of gravitational waves (GWs) is the residual displacement of test particles after the passage of the wave. GWs emitted by astrophysical sources give a memory effect and this phenomenon has been the subject of intense studies starting from \cite{Zeldovich:1974gvh} (for a review, see \cite{Favata:2010zu}). 
The memory can be measured by present or planned GW experiments, see for instance \cite{Lasky:2016knh,Nichols:2017rqr,Inchauspe:2024ibs}. Theoretically, gravitational memory is a crucial feature of the infrared (IR) limit of gravity \cite{Strominger:2014pwa} and, as such, it is deeply connected with the asymptotic symmetries of Minkowski space and soft theorems for the emission of low-energy gravitons
 (for a review of the subject see Strominger's lectures \cite{Strominger:2017zoo}).

Here we point out that the same effect exists for cosmological GWs generated in the early Universe. During inflation, GWs are produced from quantum vacuum fluctuations.  The homogeneous expanding Universe is described by a Friedmann-Lema\^itre-Robertson-Walker (FLRW)  metric, $ds^2 = a^2(\eta) \left(-d \eta^2 +  d \vec x^2 \right) $, where $\eta$ is the conformal time and $x^i$ are comoving spatial coordinates. The cosmic expansion rate is set by the Hubble parameter $H \equiv \frac{1}{a^2} \frac{d a}{d \eta} $.
Consider a  Fourier mode of physical wavenumber  $k_{\rm phys} (\eta) = k/a(\eta)$, where $k$ is a constant {\em comoving} wavenumber. When $k_{\rm phys}$ is  stretched out of the horizon by the accelerated expansion, i.e., for $k_{\rm phys} \ll H $, 
GWs freeze to a constant value.
In this regime they are unobservable because they locally correspond to an anisotropic rescaling of the coordinates. Later on, during the decelerated expansion of the Universe, they come back into the horizon, $k_{\rm phys} \gtrsim H$, and oscillate.  In this regime they affect cosmological observables, most notably the $B$-mode polarisation of the Cosmic Microwave Background (CMB) anisotropies \cite{Seljak:1996gy,Kamionkowski:1996zd}, 
which is the target of a massive observational program; see e.g.~\cite{BICEP:2021xfz,CMB-S4:2022ght,LiteBIRD:2020khw}.
As the Universe expands, the amplitude of their oscillations  decays and they disappear at late times. One can thus ask the question: is there a residual displacement of test particles that were initially at rest before the GWs came back into the horizon? Is there a memory of the primordial GWs?

\begin{figure}[t!]
\begin{center}
		\includegraphics[width=10cm]{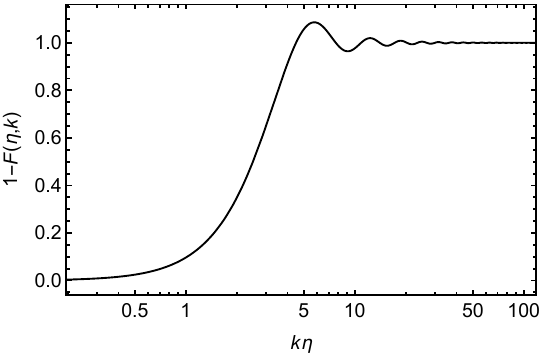}
		\caption{\small\label{fig:plotGW}Time evolution of a Fourier mode of  cosmological GWs. Here we assume that the Universe is matter dominated, in which case $F(\eta, k) = 3 [ \sin ( k \eta)/(k \eta)-  \cos(k \eta)]/(k \eta)^2$, with $k = |\vec k|$; however, the evolution is qualitatively the same for other decelerating cosmologies. The offset between early and late times is responsible of the memory effect. Notice how this is much larger than the late-time oscillations.} 
\end{center}		
\end{figure}

The calculation to the answer is very simple using the usual transverse-traceless (TT) gauge for GWs in a FLRW background. The metric in this case reads
\be
\label{FLRW}
ds^2 = a^2(\eta) \left[-d \eta^2 + (\delta_{ij} + h_{ij}(\eta, \vec x)) dx^i dx^j \right]\;,
\ee
where the tensor perturbation $h_{ij}$ is transverse, $\partial_i h_{ij} =0$, and traceless, $h^i_i=0$.  At linear order in $h_{ij} (\eta, \vec x)$,  each of its  Fourier modes are independent and we can focus on a single one, evolving as $\tilde h_{ij}(\eta, \vec k) = \bar h_{ij} (\vec k) F(\eta, k)$, where $F(\eta, k) $ is the transfer function  and $\bar h_{ij} (\vec k)$ the initial condition at early times, i.e.~for $k \ll a H$ or, equivalently, $k \eta \ll 1$. The time evolution of a mode depends on the cosmological evolution $a(\eta)$ during the decelerated expansion; however, in general one has $F \to 1$ at early times and $F \to 0$ at late times (see Fig.~\ref{fig:plotGW}). 

\begin{figure}[t!]
\begin{center}
		\includegraphics[width=7cm]{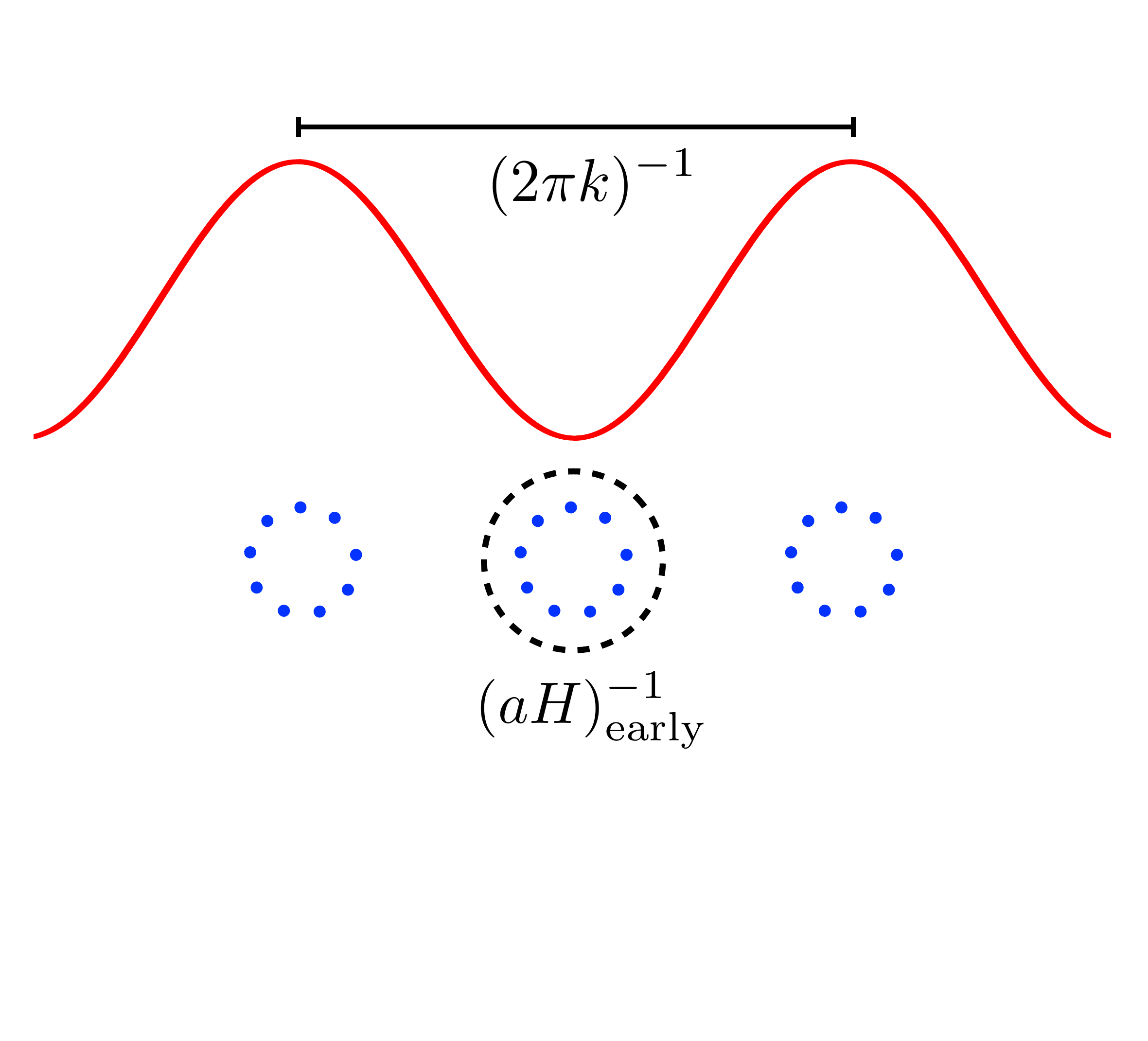}
		\hspace{1.2cm}
		\includegraphics[width=7cm]{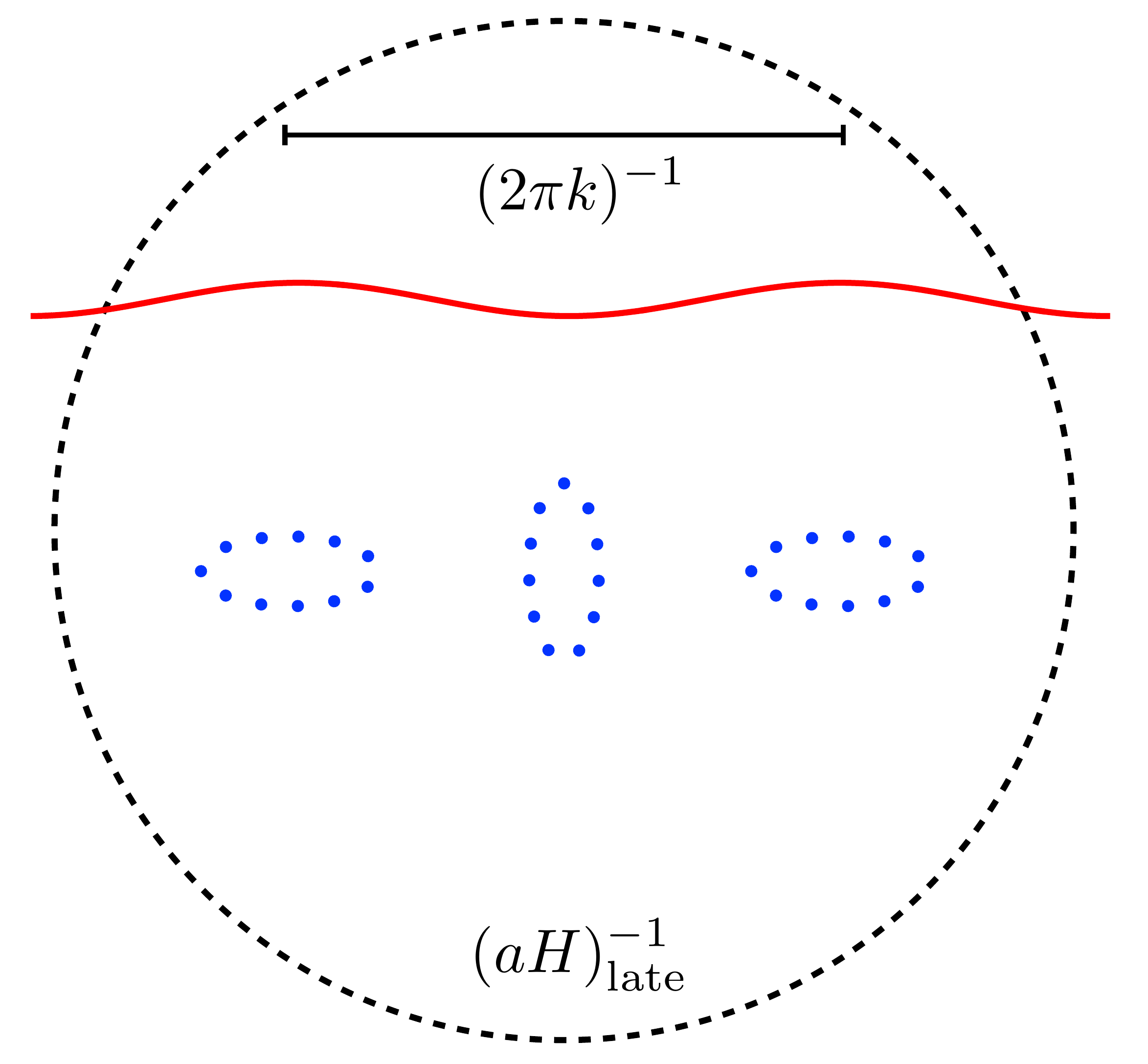}
		\caption{\label{fig:ellipses}\small At early times (left figure), the GW is much longer than the comoving Hubble horizon $(a H)_{\rm early}^{-1}$, i.e.~$k \ll (a H)_{\rm early}$, and test particles are distributed in a spherical shell. At late times (right), the GW is much shorter that the horizon, $k \gg (a H)_{\rm late}$, and has decayed away. The shell of particles is left permanently deformed to an ellipsoid.} 
\end{center}		
\end{figure}

Notice that only the spatial part of the metric is perturbed. This implies, in the same way as in the treatment of GWs around Minkowski spacetime, that in this gauge test particles that are initially at rest remain at constant spatial coordinates. Indeed, the geodesic equation reads
\be
\frac{d^2 x^i}{d \tau^2} = -\Gamma^i_{\mu\nu} \frac{d x^\mu}{d \tau}  \frac{d x^\nu}{d \tau}  = - \Gamma^i_{00} \frac{d \eta}{d \tau}  \frac{d \eta}{d \tau} = 0 \;,
\ee
where the second equality holds since there is no initial velocity and the third because $\Gamma^i_{00} = 0$ at linear order in $h_{ij}$. Therefore, test particles keep the same spatial coordinates in this gauge. If we now compare around a given point the {\em physical} distance $\vec \xi  $ between  test masses  at early times, when the GW is frozen outside the horizon, and much later, when it has decayed to zero, one finds (see Fig.~\ref{fig:ellipses})
\be
\xi^i_{\rm late} = \frac{a_{\rm late}}{a_{\rm early}} (\delta^i_j -\frac12 \bar h^i_j)\xi^j_{\rm early}  \;. 
\ee
{For an initial spherical shell, $\xi^i_{\rm early}$ has constant magnitude in all directions. At late times the spherical shell is deformed to an ellipsoid by the residual displacement induced by the GW.} 
This is the memory of the primordial GWs. To derive this equation we have assumed that the distance between the test particles is much smaller than the GW wavelength. 
 Notice also that even if the GW time-dependence $F(\eta, k)$, and therefore its effect on test particles and cosmological observables, depends on the cosmological history $a(\eta)$, the residual displacement does not: {\em the memory of primordial GWs does not depend on the cosmological evolution}.\footnote{The memory of primordial GWs, which is the topic of interest here, should not be confused with the modification of the usual Minkowski memory when the source and the detectors are in a FLRW universe (see \cite{Bieri:2015jwa,Tolish:2016ggo} and references therein).} As evident from figure \ref{fig:plotGW}, the memory of primordial GWs is large compared to the late-time oscillations. This is at variance with the memory of astrophysical binary sources, which is tiny compared to the oscillations in the waveform, since it is only a non-linear effect \cite{Christodoulou:1991cr}.

Of course, it is impossible to go back in the early Universe and measure the position of a given test mass  to compare it with its final one today. But if we look at many test masses, we can compare their initial and final positions {\em statistically}, assuming that they were homogeneously and isotropically distributed at early time. For instance, one can observe the effect of the memory of primordial GWs in the correlation function of the dark matter distribution.
The effect of a primordial GW on short-scale density perturbations was studied in \cite{Dai:2013kra,Schmidt:2013gwa}  (see also \cite{Masui:2010cz}), including the displacement of test particles as mapping between Lagrangian and Eulerian coordinates, although these references did not interpret the final displacement as a memory effect. We will discuss observational signatures below.

\begin{figure}[t!]
\begin{center}
		\includegraphics[width=10cm]{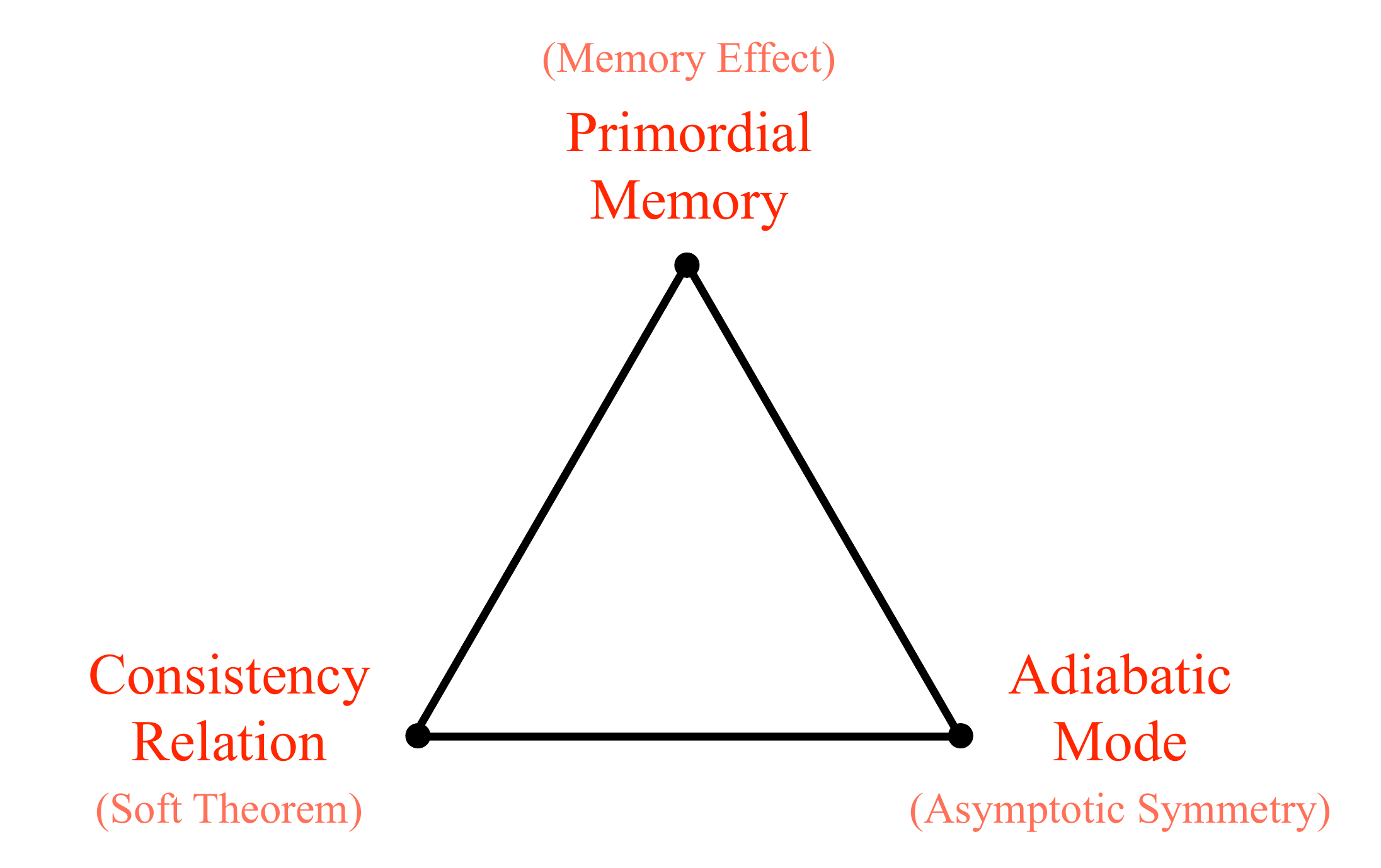}
		\caption{\small\label{fig:triangle} Cosmological Infrared Triangle: Triangle of relations in the cosmological setting (with corners Primordial Memory, Consistency Relation, Adiabatic Mode), analogue of the Infrared Triangle (with corners Memory Effect, Soft Theorem, Asymptotic Symmetry) discussed in flat space \cite{Strominger:2017zoo}.} 
\end{center}		
\end{figure}

In flat spacetime, the memory of GWs is connected to two other aspects of gravity in the IR limit: the asymptotic Bondi-Metzner-Sachs (BMS) symmetries \cite{Bondi:1962px,Sachs:1962wk} and the soft theorems for GW emission \cite{Weinberg:1965nx}. These three concepts  and their relative connection can be generalized to other massless  theories and they have been the object of a fervent research activity, with the hope of obtaining a deeper understanding of gauge and gravitational theories, see e.g.~\cite{McLoughlin:2022ljp} for a recent review. In the gravitational case, BMS symmetries, soft theorems and memory effects
form a ``triangle", as discussed in \cite{Strominger:2017zoo}. Here we point out that, similarly, the memory of primordial GWs is part of an analogous triangle, shown in Fig.~\ref{fig:triangle}, as explained below (see also \cite{Mirbabayi:2016xvc}).

Both at early times, when the GW is frozen outside the horizon, and at late times, after it has decayed away, the wave has no measurable effect and the metric is the same as an unperturbed FLRW one. However, the two unperturbed metrics are written in a different way with a diffeomorphism relating the two,
\be\label{eq:diff}
x_{\rm late}^ i = \left(\delta^i_j + \frac12 \bar h^i_j \right)x^j_{\rm early} \;.
\ee
The memory effect is a measure of this ``mismatch" of the coordinates. This is in complete analogy with what happens in flat space. The flat space before and after the passage of the GW are related by a diffeomorphism, a BMS transformation, and the memory effect is a measure of this mapping.  Cosmological perturbations that reduce to a pure diffeomorphism in the long-wavelength limit are called adiabatic \cite{Weinberg:2003sw}. Adiabaticity has important implications: for instance it implies that $ h_{ij}$ is constant in time, independently of the evolution of the Universe. Notice that the memory effect does not depend on the production mechanism (classical or quantum) of GWs during inflation. What matters is that GWs get frozen out of the cosmological horizon during the accelerated expansion of the Universe and become adiabatic.\footnote{A memory effect associated with the quantum production of the GWs {\em during} inflation has been studied in \cite{Ferreira:2016hee}, although it is not an observable quantity.}

Adiabatic modes are related to the third corner of the triangle: the consistency relations of cosmological perturbations. These originate from the adiabaticity of the modes {\em during} inflation. Let us imagine we want to calculate the correlation between a long GW (with polarisation $s$ and polarisation tensor $\epsilon_{ij}^s$) and short scalar perturbations described by the variable $\zeta$. The long tensor mode is already frozen outside the horizon during inflation when the short scalar modes are generated. Therefore there is no physical effect of the GW apart from the diffeomorphism of eq.~\eqref{eq:diff}. This logic gives a universal relation in this limit of the form \cite{Maldacena:2002vr} 
\be\label{consGW}
\langle h^s_{\vec k_1} \zeta_{\vec k_2} \zeta_{\vec k_3} \rangle' \simeq - \langle h^s_{\vec k_1} h^s_{-\vec k_1} \rangle' \epsilon_{ij}^s k_2^i k_2^j \frac{\partial}{\partial k_2^2} \langle\zeta_{\vec k_3} \zeta_{\vec k_2} \rangle' \;,
\ee
where the prime denotes that we have stripped the momentum conserving delta function off the statistical averages $\langle \cdots \rangle$. 
This relation shows the effect in Fourier space of the anisotropic rescaling of the coordinates.\footnote{Consistency relations and adiabatic modes exist also for scalar perturbations. However, since a scalar perturbation does not decay away when entering the horizon, one cannot define a memory effect.} 

Not only do consistency relations  apply during inflation, but they can be extended to observable quantities by including the late-time effect of the long mode on the short ones, see for instance \cite{Creminelli:2011sq, Huang:2012ub, Peloso:2013zw, Creminelli:2013mca}. To show the link between consistency relations and memory, let us consider for example the baryon acoustic oscillations (BAOs), i.e.~the fluctuations in the density of matter due to acoustic waves in the primordial plasma. The BAO peak can be crudely represented as a thin spherical shell of radius $r_{\rm BAO}$ around a given galaxy, where it is more likely to find another galaxy. (Notice the analogy with the spherical shell of inertial detectors of Fig.~\ref{fig:ellipses}.) What happens to the BAO shell in the presence of a GW with wavelength much longer than $r_{\rm BAO}$? Since the scalar modes giving rise to the BAOs are much shorter than the GW, they are generated during inflation when the GW is already frozen outside the horizon. The BAO is thus a sphere with constant physical radius when the GW is outside the horizon. When the GW comes back and fade away, the BAO shell will be distorted to an ellipsoid due to the memory effect. This represents a physical effect associated with the consistency relation eq.~\eqref{consGW}.

The effect of anisotropic distortion due to the GW on the late-time dark matter density contrast   $\delta$ can be computed rescaling the  coordinates using eq.~\eqref{eq:diff} and
reads
\cite{Schmidt:2013gwa} 
\be
\label{SPZ}
\delta  (\eta, \vec x) |_h = \left( 1 +  \frac12 \bar h_{ij}  x^i \partial^j  \right) \delta  (\eta, \vec x) \;.
\ee
The same formula applies to  the galaxy number over-density or to the overdensity of other cosmological tracers. 
The GW modifies the small-scale correlation  between  dark matter particles or galaxies. This effect can be compared to that of the gravitational tidal field caused by dark matter itself, responsible for the  deformation of the BAO sphere. It is obtained by replacing $\bar h_{ij}$ in the above equation by  $\partial_i \partial_j \nabla^{-2} \delta$: this effect is therefore much larger than the one  of the memory of primordial GWs.

Let us comment on the possibility of measuring the memory of primordial GWs in galaxy surveys by observing its effect on the alignment of the shapes of the observed galaxies: the so-called intrinsic alignment.\footnote{The second source of alignment is the small deflection of light due to foreground dark matter distribution: the weak lensing. Light is also deflected by the presence of GWs, although this is a very tiny effect \cite{Dodelson:2003bv,Schmidt:2012nw}.}  Although small, this effect can be measured in the $B$-mode of the weak lensing shear. The memory of the primordial GWs  contributes to the intrinsic  alignment, since the matter that will give rise to a certain galaxy is displaced following eq.~\eqref{SPZ}.   Going beyond the test particle limit and taking into account the gravitational interaction of matter, GWs change the growth on short scales: this tidal interaction is an effect of the same order as the memory, but with a different scale dependence and with an amplitude that depends on the cosmological evolution \cite{Dai:2013kra,Schmidt:2013gwa,Akitsu:2022lkl}. It would be nice to study the separability of the two effects on the intrinsic alignment or to explore the possibility of detecting the primordial memory effect using other cosmological probes.

Notably, the polarization of the CMB is affected by the primordial memory. Indeed, as we discussed, the most direct way of detecting primordial GWs is by observing  $B$-modes of the CMB. Their spectrum can be read from the book of Gorbunov and Rubakov \cite{Gorbunov:2011zzc} and is given by [see eq.~(10.83) of that reference]
\be
C_\ell^{BB} = \frac{4 \pi}{25} \Delta \eta_r^2 \int_0^{\infty} \frac{d k}{k} {\cal P}_T (k) \left|  F ' ( k , \eta)  \right|_{\eta=\eta_r}^2 \left[ \frac{\ell+2}{2 \ell+1} j_{\ell -1} (k \eta_0) - \frac{\ell - 1}{2 \ell+1} j_{\ell +1} (k \eta_0)  \right]^2 \;,
\ee
where $\Delta \eta_r$ is the thickness of recombination surface,  ${\cal P}_T (k)$ is the power spectrum of the primordial tensor modes $\bar h_{ij}$ and $F' \equiv \partial F/\partial \eta$.
This observable is  sensitive to the derivative of $F$ with respect to $\eta$ evaluated at recombination,  $\eta=\eta_r$. Therefore, it is not directly sensitive to the memory of primordial gravitational waves. However, with sufficient precision this observation can still be used to reconstruct the variation of $F$ from early to late time and measure the memory of the waves. Indeed, while $C_\ell^{BB}$ depends   on $F'$ at $\eta_r$, its $\eta$-dependence can be reconstructed from its $k$-dependence. For instance, in matter domination 
$F$ is a function of the combination $k \eta$.\footnote{Since the power spectrum is sensitive to $|F'|^2$, and not directly to $F'$, there is a sign ``ambiguity'' in reconstructing the  derivative.} As a matter of fact, since the first peak of the $B$-mode power spectrum is much larger than the others, one can see ``by eye" that the wave is not oscillating around zero. Notice that the detection of the primordial memory requires to be able to compare moments when the GW is outside and inside the horizon. This is clearly not possible for interferometers, even in inflationary models with a detectable GW spectrum.

A new kind of gravitational memory in Minkowski spacetime, dubbed ``Spin Memory'', has been proposed \cite{Pasterski:2015tva}, consisting in the relative time delay acquired by a Sagnac interferometer  after the passage of a GW. See also \cite{Flanagan:2018yzh,Seraj:2021rxd, Seraj:2022qyt}. In analogy to the standard gravitational memory, this effect is deeply connected to extensions of the asymptotic symmetries of Minkowski spacetime \cite{Barnich:2009se} and to soft-graviton theorems at {\em subleading} order in the soft expansion \cite{Cachazo:2014fwa}. 

Also the primordial memory can be extended to subleading order. Imagine that the test particles previously considered are replaced by gyroscopes. These gyroscopes are initially aligned when the GW is well outside the Hubble radius. We want to show that the gyroscopes are not aligned anymore after the GW has decayed away and that this memory is sensitive to the derivative of the initial value of the wave, $\partial_k \bar h_{ij}$. The gyroscopes are described by space-like 4-vectors, $S^\nu$, which are parallel-transported along time-like geodesics with tangent vector $V^\mu$. This gives
\be
0 = V^\mu \nabla_\mu S^i = \partial_0 S^i + \Gamma^i_{0 \alpha} S^\alpha = \dot S^i + \frac12 \dot h_{ij} S^j  \;.
\ee
In the second equality we specialized to the geodesics that are initially at rest: in the coordinates we are using their tangent vector is simply $V^\mu = (1,\vec 0)$. In the third equality we used that the 4-vectors $V^\mu$ and $S^\mu$ are orthogonal, since they are orthogonal at the beginning and then both parallelly transported. We have also evaluated the Christoffel's symbol in our gauge, eq.~\eqref{FLRW}. 

This equation does not describe any rotation of the gyroscopes, but simply a rescaling, which is necessary to keep the norm of the vector constant. For simplicity, let us assume that $\bar h_{ij}$ vanishes at the position of the experiment. The equation above then says that the spatial derivatives $\partial_k \bar h_{ij}$ do not induce a change of $S^i$: the vectors $S^i$ remain constant in this gauge, like the coordinates of the geodesics in the previous discussion. However this does not mean there is no physical rotation of the gyroscopes. At early times, in order to start with parallel gyroscopes, one has to perform a parallel transport on a surface of constant time\footnote{Since we are only considering first derivatives of the metric, the space is flat and the parallel transport does not depend on the path.}. To move in a spatial direction one has to parallel transport the vector $S^i$ along a spatial vector $W^j$, i.e.~solve $W^j (\partial_j S^i + \Gamma^i_{jk} S^k) = 0$. This gives the expression of parallel gyroscopes around a central one $S^i_{(0)}$
\be
S^i = S^i_{(0)} +x^j (-\Gamma^i_{jk}) S^k_{(0)} \;,
\ee
at first order in $x^i$.
When the GW has decayed away, the spatial metric is just $\delta_{ij}$ and the gyroscopes are clearly not aligned. Writing the Christoffel's symbols in terms of $\bar h_{ij}$, one gets
\be
\Delta S^i = -\frac12 (\partial_j \bar h_{ik}+ \partial_k \bar h_{ij}- \partial_i \bar h_{jk}) x^j S^k \;.
\ee
The memory of the direction of the gyroscopes is encoded in the spatial derivative of the GW. 

Adiabatic modes and consistency relations were studied at subleading order in the soft limit, see for instance \cite{Creminelli:2012ed,Hinterbichler:2013dpa}: therefore the whole triangle exists also at subleading order, in analogy with the Minkowski case.

In the Minkowski case, besides the displacement memory, there exists also a {\em velocity} memory, i.e.~a change in the relative velocity of two test masses (see for instance \cite{Bieri:2024ios} and references therein). It quite easy to study this effect is the case of primordial GWs\footnote{We thank the anonymous referee for raising this question.}. At early times, when the GW stays constant outside the horizon, let us consider a set of test particles with relative velocities with respect to a comoving observer. We can take the test particles to have the same magnitude of relative velocity $|v_{\rm early}^i|$, but different directions: we can imagine them as flying away from the comoving observer in all directions in an isotropic way. Eventually the GW will come back in the horizon and decay away as discussed above. The question is what happens to the relative velocities with respect to the comoving observer. Does the GW induce an anisotropy in these velocities that we assumed isotropic at early times? At an initial time all particles have the same physical velocity with respect to a comoving observer. This means that
\be
|v_{\rm early}^i| = \sqrt{a_{\rm early}^2 \cdot (\delta_{ij}+ \bar h_{ij})\frac{dx^i}{d\tau} \frac{dx^j}{d\tau} }
\ee
is the same for all the particles flying away from the comoving observer. In the limit in which one can neglect the spacial variation of $h_{ij}(\eta)$, the metric is invariant under spacial translations and along a geodesic the following quantity remains constant
\be
a^2 \cdot (\delta_{ij}+ h_{ij}(\eta)) \frac{dx^i}{d\tau} = {\rm const.}
\ee
Using this conservation law, one can deduce that at late times, after the GW has decayed to zero, the final velocities of the test particles are
\be
v_{\rm late}^i = v_{\rm early}^j \frac{a_{\rm early}}{a_{\rm late}} (\delta_j^i + \frac12 \bar h_j^i) \;.
\ee
The norm of the velocity $v_{\rm late}$ now depends on the direction: the anisotropic expansion due to the GW induces a different redshift of the peculiar velocities in different directions. Notice that, as in the case of the displacement memory, the anisotropy just depends on the asymptotic value $\bar h_{ij}$ and is not sensitive to the cosmological history. This change in velocities might be relevant for redshift space distortions.

\vspace{0.3cm}
{\bf Acknowledgements:} FV acknowledges partial support by the ANR Project COLSS (ANR-21-CE31-0029) and CNES. We thank Laura Donnay, Marc Favata, Luca Santoni, Fabian Schmidt, Andy Strominger and Sasha Zhiboedov for useful discussions. We thank Fabian Schmidt for comments on the draft.

\footnotesize

\bibliographystyle{utphys}
\bibliography{ref}
\end{document}